\documentclass[10pt,a4paper]{article}
\usepackage[T1]{fontenc}
\usepackage[utf8]{inputenc}
\usepackage{authblk}
\usepackage{amsmath}
\usepackage{amsfonts}
\usepackage{amssymb}
\usepackage{graphicx}
\usepackage{multirow}

\title{Anisotropic metamaterial optical fibers}
\author[1]{Dheeraj Pratap\thanks{pdheeraj@iitk.ac.in}}
\author[1]{S. Anantha Ramakrishna}
\author[2]{Justin G. Pollock}
\author[2]{Ashwin K. Iyer}
\affil[1]{Department of Physics, Indian Institute of Technology, Kanpur 208016, India}
\affil[2]{Dept. of Electrical and Computer Engineering, University of Alberta Edmonton, Alberta T6G 2V4 Canada}
\date{}
\begin{document}
\maketitle
\begin{abstract}
Internal physical structure can drastically modify the properties of waveguides: photonic crystal fibers are able to confine light inside a hollow air core by Bragg scattering from a periodic array of holes, while metamaterial loaded waveguides for microwaves can support propagation at frequencies well below cutoff.  Anisotropic metamaterials assembled into cylindrically symmetric geometries constitute light-guiding structures that support new kinds of exotic modes.  A microtube of anodized nanoporous  alumina, with nanopores radially emanating from the inner wall to the outer surface, is  a manifestation of such an anisotropic metamaterial optical fiber. The nanopores, when filled with a plasmonic metal such as silver or gold, greatly increase the electromagnetic anisotropy. The modal solutions in  such anisotropic circular waveguides can be uncommon Bessel functions with imaginary orders. 
\end{abstract}
Optical  fibers form the backbone of optical communications systems worldwide. Their analogues in the microwave regime are hollow or coaxial metallic waveguides, which are indispensable in applications requiring field confinement, low losses, and high power-handling capability. The introduction of physical structure in the form of a periodic array of microscopic air holes running along the fiber axis in the photonic crystal fiber~\cite{Russell} gives rise to new possibilities. Light can even be confined within a hollow core or a core of lower refractive index due to confinement caused by a photonic band gap in the surrounding region containing a periodic array of holes.  Alternatively for a solid core surrounded by microholes, the clad region with the microholes effectively creates a region with lower modal index for light propagating in the core and confines light by a modified total-internal-reflection effect.  Birefringence of the modes in these fibers usually results from a two-fold asymmetry, and  even small birefringence created by applied strain provides for many sensor applications~\cite{PCFcoil}. However, these fibers usually have no structure or structural anisotropy perpendicular to the fiber axis.  

Metamaterials~\cite{sarbook} are structured composite materials that can be designed for specific electromagnetic responses across the spectrum, not readily available in nature.  The structural units of the metamaterials are  subwavelength in size, which makes metamaterials amenable to description as effective media. The geometry and materials of the  structural units can be chosen to yield a wide variety of  desired responses and dispersions for the metamaterial. Structural anisotropy of the units results in anisotropic effective-medium parameters, such as the dielectric permittivity or magnetic permeability. This anisotropy can be extremely large compared to natural anisotropic responses in crystals. For example, plasmonic nanowires embedded in a nanoporous dielectric medium such as nanoporous alumina are known to have negative permittivity along the nanowire's axes while having positive permittivities in the orthogonal directions~\cite{sridhar_APL}.

Interesting phenomena have been shown in microwave rectangular and circular waveguides loaded with metamaterials, including propagation at frequencies below the fundamental mode of the unfilled waveguide~\cite{unmu,belowcutoff}.  Similarly, the unprecedented control over the metamaterial properties as well as the enormous flexibility of applications possible with a fiber makes it attractive to combine them into a common platform.  Theoretical discussions of fibers made of negative-refractive-index materials have indicated interesting properties such as sign-varying energy flux~\cite{novitsky_barkovsky} and zero group-velocity dispersion~\cite{novitsky}. Surface waves guided along a cylindrical metamaterial waveguide were discussed in~\cite{cory_blum}.  Dispersion of modes in a fiber with anisotropic dielectric permittivity was also discussed recently~\cite{IIT_kgp}. However, the immense difficulty of practically  assembling small nanostructured metamaterial units inside the micrometer sized fibers has essentially discouraged the discussion of such systems for optical frequencies.  Only some cylindrically layered tubes, or those with coaxially oriented nanowires that are otherwise uniform along the axis, have been theoretically discussed~\cite{Smith,emode} as metamaterial optical fibers.  Recently, guided modes in a hollow core waveguide with a uniaxial metamaterial cladding have been theoretically discussed~\cite{atakaramians_JOSAB_2012}, where the metamaterial was proposed to to consist of a layer of drawn split-ring resonators. The dispersion of the transverse components of the magnetic and dielectric properties of the metamaterial clad was found to dominate the behavior of the modes~\cite{atakaramians_JOSAB_2013} and a drawn fiber with a single split-ring resonator at the fiber axis was demonstrated for terahertz frequencies in~\cite{nsingh_OE2012}.  We note that all the proposals for metamaterial fibers so far have considered only designs that are invariant and homogeneous along the fiber axis, i.e., effectively the problem is only two-dimensional, presumably because only the drawing technique was considered for the fabrication of these fibers. This limits the discussion to a form of anisotropy with effective dielectric parameters $\varepsilon_r \simeq \varepsilon_\phi \ne \varepsilon_z$. 

We demonstrate here that anodization techniques~\cite{Masuda} with aluminum micro-wires can result in a three-dimensionally structured metamaterial fiber made of nanoporous alumina (Al$_2$O$_3$).  Such a fiber may be regarded to be a type of circular waveguide containing an anisotropic metamaterial, and embodies a very unique system that can support highly uncommon modes described by Bessel and Neumann functions of imaginary orders.  A nanoporous alumina microtube with nanopores that emanates radially from an inner surface upto a nanoporous surface provides a physical manifestation of an anisotropic  metamaterial fiber where the limitation $\varepsilon_r = \varepsilon_\phi$ is lifted. Further, the true nanometric sizes of the structure involved here makes a homogenized description of the metamaterial structure more valid. The nanopore size and density, which determine the effective-medium properties of the ansiotropic metamaterial, can be flexibly controlled using the initial size of the nanowire and the anodization  process parameters.   The nanopores can also be filled up by various materials using well-known methods and makes this system particularly attractive for sensor applications.  The possibility of filling up the nanopores with plasmonic metals like gold and silver makes possible a fiber made of a hyperbolic metamaterial with large resonant plasmonic interactions  between the  metallic inclusions in the nanopores.  Furthermore, the central core of the metamaterial fiber could consist of a metallic core, a hollow core, or a core filled up by various other materials, thereby  yielding a system that can be flexibly designed for various processes.  Thus, the metamaterial could form the core itself or a clad that confines the guided energy within a homogenous core. 
\begin{figure}
\begin{center}
\includegraphics[width=110mm]{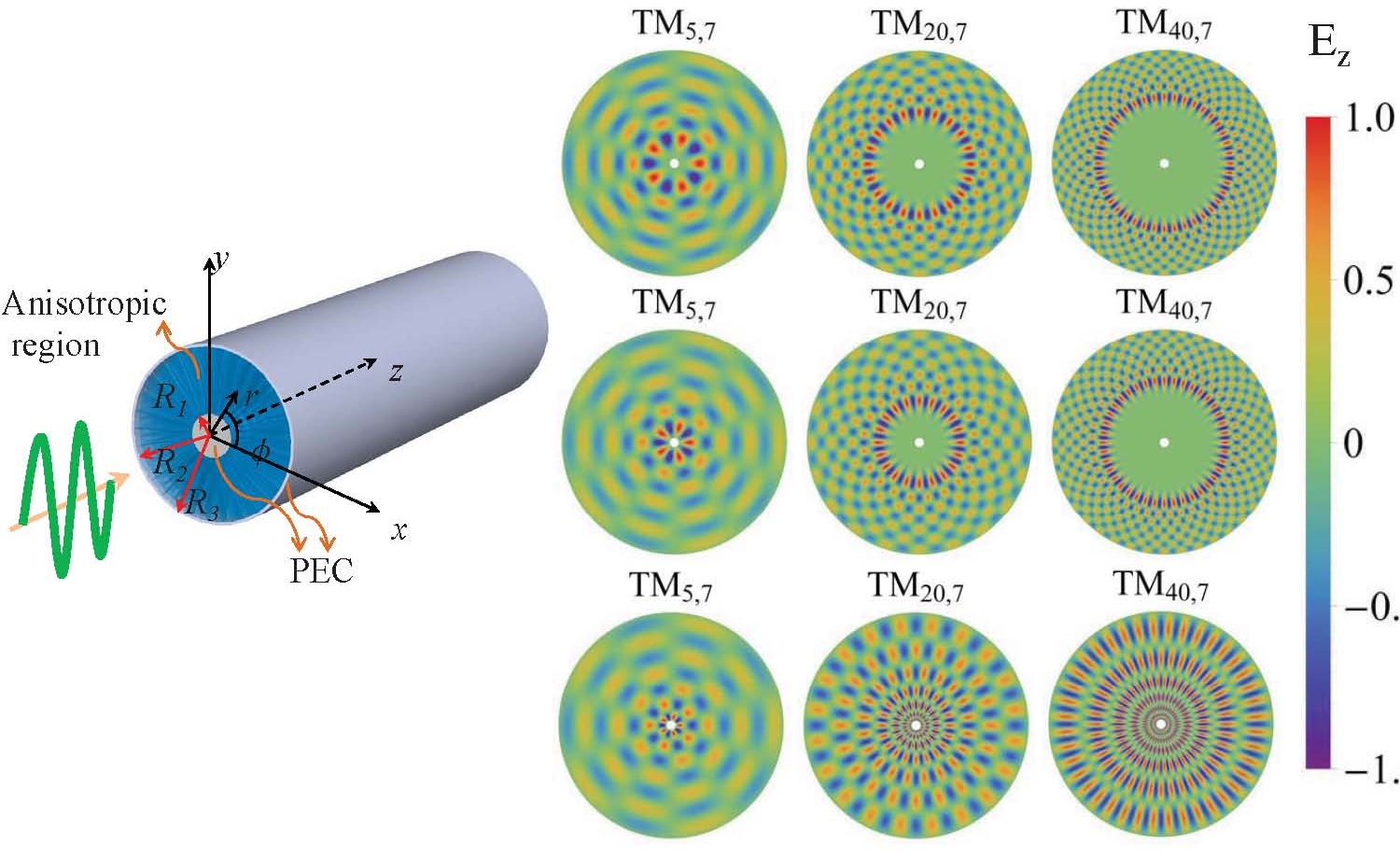}
\end{center}
\caption{ \label{fg:schematic} Left: A schematic of the anisotropic optical fiber with PEC core and outer boundary. Right: panels show the computed electric fields ($E_z$) of modes for $R_1=0.5~\mu$m, $R_2=12.5~\mu$m with $m=5$ (left column), $m=20$ (middle column) and $m=40$ (right column). Top row shows the Bessel modes in an isotropic alumina fiber with real integral orders for $\varepsilon_r = \varepsilon_\phi = \varepsilon_z= 3.118$. The middle row shows Bessel modes  in an anisotropic nanoporous alumina fiber with fractional order and positive dielectric permittivity components for $\varepsilon_r = 2.467 $ and $\varepsilon_\phi =\varepsilon_z= 2.638$. The bottom row shows the fields for Bessel modes of anisotropic nanoporous alumina fiber with imaginary orders and $\varepsilon_r = 2.638$ and $\varepsilon_\phi =\varepsilon_z= 2.467 $. Relative permittivity components are at wavelength 633 nm.}
\end{figure}

We start by considering a cylindrically symmetric waveguide formed from a material with relative magnetic permeability $\mu = 1$ and relative dielectric permittivity tensor as 
\begin{equation}
\bar{\bar{\varepsilon}} = \varepsilon_0 \left(\begin{array}{ccc}   \varepsilon_r & 0 & 0  \\
								0 & \varepsilon_\phi & 0 \\
								 0 & 0 & \varepsilon_z \end{array} \right).
\end{equation}
That is, the waveguide material has different permittivity responses along the radial, azimuthal, and axial ($z$) directions.  To understand the nature of the modes in such  waveguide, we will assume that the material is spatially homogeneous and that the medium is lossless (the components of $\varepsilon$ are real) and dispersionless -- an assumption that will be relaxed later.  

For simplicity, we will deal with the coaxial case in which the concentric inner ($r=R_1$) and outer ($r=R_2$) surfaces of the waveguide are perfect electric conductors (PEC)(see Fig. \ref{fg:schematic} for a schematic).  This enables us to concentrate on the effects that arise due to propagation in the anisotropic medium without worrying about issues of confinement. This is not an essential requirement as will be discussed later. In this case, the transverse electric (TE) modes with $E_z=0$ and the transverse magnetic (TM) modes with $H_z =0$ decouple and the Maxwell's equations become separable. We obtain solutions within the anisotropic medium for the TE modes as
\begin{equation}
H_z = \left[ A J_\nu (k_r r)+ B Y_\nu(k_r r)\right] \exp [i(m\phi + \beta z)],
\end{equation}
where $J_\nu$ and $Y_\nu$ are the Bessel and Neumann functions of  order $\nu$, $\beta$ is the propagation constant, and $m$ is a non-zero integer. To simplify, if one considers only dielectric anisotropy with $\mu=1$, and $\varepsilon_z=\varepsilon_\phi$, we have 
\begin{equation}
k_r^2 =\varepsilon_\phi\frac{\omega^2}{c^2} - \beta^2,~~~\mathrm{and}~~~\nu^2 = \frac{(\varepsilon_\phi \omega^2/c^2 - \beta^2)}{(\varepsilon_r\omega^2/c^2 - \beta^2)}m^2.
\end{equation}
The coefficients $A$ and $B$ can be found by putting parallel component $E_\phi$ to zero at PEC boundary $R_1$ and $R_2$ and getting the following condition
\begin{equation}
J_{\nu}^{'}\left(k_{r}R_{1}\right) Y_{\nu}^{'}\left(k_{r}R_{2}\right) - J_{\nu}^{'}\left(k_{r}R_{2}\right) Y_{\nu}^{'}\left(k_{r}R_{1}\right) = 0,
\label{eqn_bd_condn}
\end{equation}
where prime ($'$) indicates the radial derivative. Analogous expressions are obtained for the TM modes as
\begin{equation}
E_z = \left[ A J_\nu (k_r r)+ B Y_\nu(k_r r)\right] \exp [i(m\phi + \beta z)],
\end{equation}
where
\begin{equation}
\label{eqn_TM_dispersion}
k_r^2 =\frac{\varepsilon_z}{\varepsilon_r}\left( \varepsilon_r\omega^2/c^2 - \beta^2 \right),~~~\mathrm{and}~~~ 
\nu^2 = \frac{\varepsilon_\phi}{\varepsilon_r}\left(\frac{\varepsilon_r\omega ^2/c^2 - \beta^2 }{\varepsilon_\phi\omega^2/c^2 - \beta^2}\right)m^2.
\end{equation}
 The coefficients $A$ and $B$ are related by setting the tangential components of the field,  $E_z = 0$ for the TM mode, at the PEC boundaries, resulting in
\begin{equation}
J_{\nu}\left(k_{r}R_{1}\right) Y_{\nu}\left(k_{r}R_{2}\right) - J_{\nu}\left(k_{r}R_{2}\right) Y_{\nu}\left(k_{r}R_{1}\right) = 0,
\label{eqn_bd_condn}
\end{equation}
which determines $k_r$ and $\beta$ and consequently, the dispersion of the mode. For a given $m$, if $n$ is the number of zeroes of the Bessel mode within the fiber radius, then in general, the given mode will have $n$ lobes in the radial direction and can be labelled by the pair of numbers $(m,n)$.
\begin{table}
\caption{Table showing the conditions on the material parameters and the propagation constant to obtain imaginary orders ($\nu$) for the Bessel functions that describe the modes for the TE and TM polarizations in the anisotropic fiber. Note that $k_0^2 = \omega^2 / c^2$. \label{table_imaginary_nu}}
\noindent\begin{tabular}[h]{l|c|c||c} \hline 
Mode & $k_r$ & {Conditions for $k_r$}   & {Requirements for $\nu^2 <0$ }\\ \hline
\multirow{2}{*}{\bf  TE} & Real & $\varepsilon_\phi > \beta^2 /k_0^{2}$ & $\varepsilon_r < \beta^2 /k_0^{2} < \varepsilon_\phi$ \\
 & Imag. &  $\varepsilon_\phi < \beta^2 /k_0^{2}$ & $\varepsilon_\phi < \beta^2 /k_0^{2} < \varepsilon_r$ \\ \hline
 \multirow{4}{*}{\bf TM} & \multirow{4}{*} {Real} &  \multirow{2}{*} {$\varepsilon_z/\varepsilon_r > 0$, $\varepsilon_r > \beta^2 /k_0^{2}$}   & {$\varepsilon_\phi/ \varepsilon_r < 0$, $\varepsilon_r > \beta^2 /k_0^{2}$, $\varepsilon_\phi > \beta^2 /k_0^{2}$} \\ \cline{4-4}
 &  &  & {$\varepsilon_\phi/\varepsilon_r > 0$, $\varepsilon_\phi < \beta^2 /k_0^{2} < \varepsilon_r$} \\ 
\cline{3-4}
  &  & \multirow{2}{*} {$\varepsilon_z/\varepsilon_r < 0$, $\varepsilon_r < \beta^2 /k_0^{2}$ } & 
  {$\varepsilon_\phi/\varepsilon_r < 0$, $\varepsilon_r < \beta^2 /k_0^{2}$, $\varepsilon_\phi < \beta^2 /k_0^{2}$} \\ \cline{4-4}
 & & &  { $\varepsilon_\phi/\varepsilon_r > 0$ and $\varepsilon_r < \beta^2 /k_0^{2} < \varepsilon_\phi$ }\\ \hline
\end{tabular} 
\end{table}

The nature of the modes critically depends on $\nu$ and $k_r$.  It is well known that when $k_r$ is imaginary, the modal solutions assume the form of the modified Bessel functions.  More importantly, in this case, the anisotropic nature of the waveguide allows the order, $\nu$, of the Bessel function to be fractional and even imaginary ($\nu^2<0$)~\cite{Dunster,Chapman}, which is not possible in isotropic systems.  Consider, for example, a  propagating ($\beta$ is real) TE mode  with with real $k_r$ (implying that $\varepsilon_\phi > \beta^2 c^2 /\omega^2$). If $\varepsilon_r < \beta^2 c^2 /\omega^2 < \varepsilon_\phi$, then $\nu^2 <0$ and the order of the mode becomes imaginary.  This is seen to occur straightforwardly in a medium with $\varepsilon_r<0$ and $\varepsilon_\phi>0$, whereas in a medium with $\varepsilon_r = \varepsilon_\phi$, this situation would not arise. Note that the inequalities reverse if we seek a TE mode with imaginary $k_r$ described by the modified Bessel functions.  The requirements for an imaginary order for the TM modes are slightly more involved due to the various possibilities on the material parameters, and Table~\ref{table_imaginary_nu} encapsulates these conditions for $\nu^2 <0$ with $k_0^2 = \omega^2 / c^2$.  Note that some simplifications occur when $\varepsilon_z= \varepsilon_\phi$ which will be approximately true for our physical implementation. 

\begin{figure}
\begin{center}
\includegraphics[width=80mm]{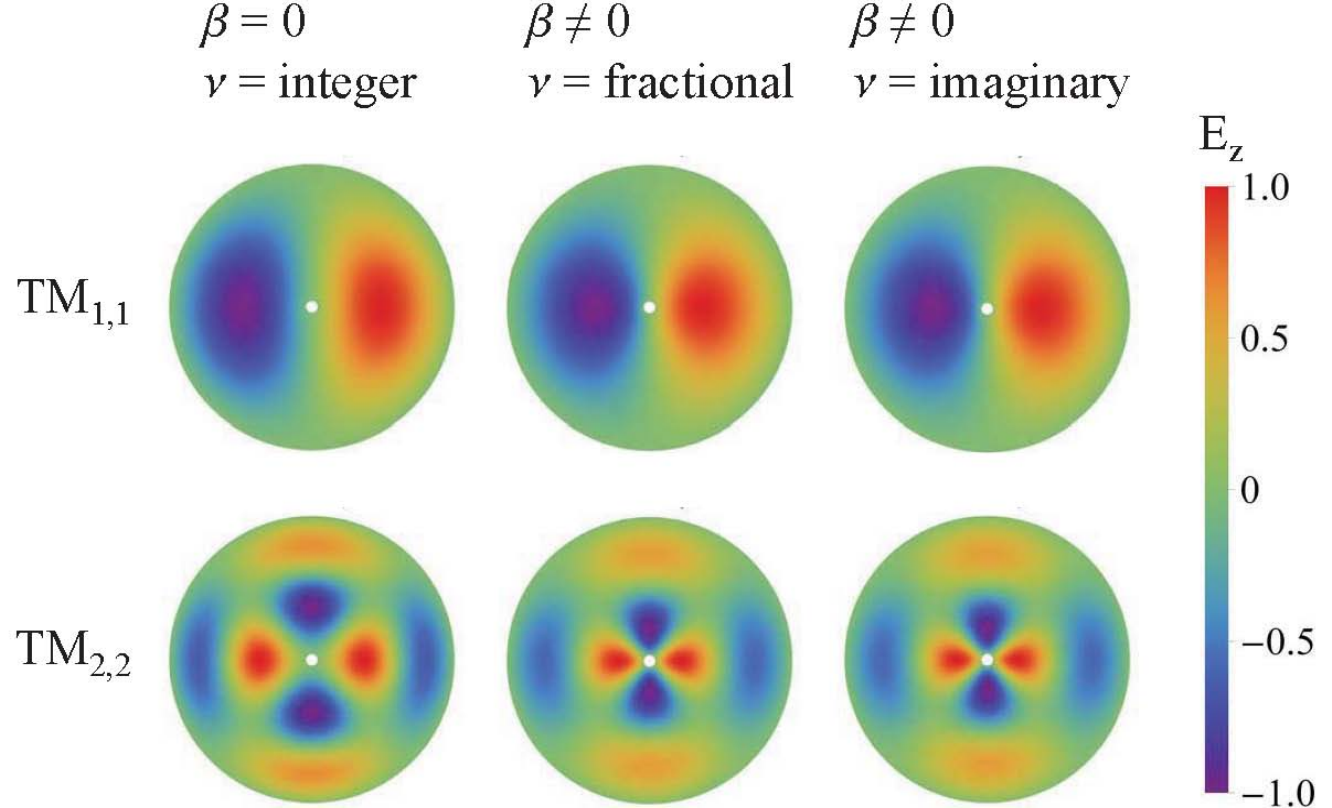}
\end{center}
\caption{ \label{fg:lower_modes} The behavior of low-order modes (TM$_{1,1}$ and TM$_{2,2}$) in a anisotropic nanoporous alumina fiber for dimensions $R_1=0.5~\mu$m and $R_2=12.5~\mu$m. Col. 1: Eigen modes at $\beta=0$, $\nu=m$ integral order, Col. 2: modes at $\varepsilon_r = 2.467 $, $\varepsilon_\phi =\varepsilon_z= 2.638$, and $\beta \neq 0$, $\nu \neq m$ fractional order, Col. 3: modes at $\varepsilon_r = 2.638$, $\varepsilon_\phi =\varepsilon_z= 2.467 $, and $\beta \neq 0$, $\nu \neq m$ imaginary order. Dielectric constants are at wavelength 633 nm.}
\end{figure}

Physical applications of Bessel functions with imaginary orders have been rarely reported~\cite{Dunster,Chapman,Grimshaw}.
There are consequences to a fractional or imaginary order, $\nu$.  The order of the Bessel function and the field distributions now depend on the propagation constant $\beta$, unlike those for an integral order ($m$) mode in isotropic media. Thus, implementing the boundary conditions in  Eq. (\ref{eqn_bd_condn}) becomes more involved. To verify our results, full-wave simulations on a cross-section of the anisotropic fiber were performed using the COMSOL® software suite. A comparison of the cut-off frequencies calculated for some modes by the software and the analytic result are given in Table~\ref{table_cutoffs}. While at cut-off, the mode has an integral order, the order $\nu$ of the Bessel function changes with $\beta$ or the frequency, $\omega$, for a given $m$.  One of the consequences is that the modes become comparatively more confined near the centre with increasing $\beta$ or $\omega$. This confinement is more drastic than in ordinary dielectric fibers as the effects of $\beta$ or $\omega$ are manifested within the Bessel functions’ order ($\nu$) rather than only the argument ($k_r$).  Fig.~\ref{fg:lower_modes} shows this behavior for the TM$_{1,1}$ and TM $_{2,2}$ modes. In contrast to a waveguide filled with an isotropic medium where the argument of the Bessel function ($k_r$) changes with $\beta$, here, the order $\nu$ of the Bessel function itself evolves with $\beta$.
 \begin{table} \begin{center}
 \caption{Table showing the cutoff frequencies calculated for an anisotropic (coaxial) fiber of outer diameter 25 $\mu$m and inner diameter of 1 $\mu$m and the waveguide is homogeneously filled with a material of dielectric permittivity of $\varepsilon_r = 2.638$ and $\varepsilon_\phi =\varepsilon_z= 2.467 $. \label{table_cutoffs}}
\begin{tabular}[h]{lcc} \hline 
\multirow{2}{*} {Mode} &  \multicolumn{2}{c} {Cutoff Frequency (THz) }\\
& {Analytic} & {COMSOL\textregistered} \\ \hline
TM$_{0,1}$ & 	 7.319   &     7.314   \\
TM$_{0,2}$ &    15.405   &    15.395   \\
TM$_{1,1}$ &     9.363   &     9.356   \\
TM$_{1,2}$ &    17.202   &    17.190   \\
TM$_{2,1}$ &    12.489   &    12.481   \\
TM$_{2,2}$ &    20.472   &    20.458   \\
TM$_{5,5}$ &    54.029   &    54.058   \\
TM$_{10,5}$ &   70.248   &    70.223   \\ \hline 
\end{tabular} \end{center}
\end{table}

 The modes with imaginary orders are very interesting in that the higher-order (large $m$) modes are localized near the central region of the waveguide (see Fig.~\ref{fg:schematic}). In contrast, the large $m$ whispering gallery modes are concentrated at the edges of a waveguide filled with an isotropic material.  Further, these  Bessel functions with imaginary orders undergo large oscillations near the center and need not converge at $r=0$~\cite{Dunster} -- a reflection of the geometric singularity at the axis, which cannot be attained in a physical system. The PEC  boundary condition applied at $r=R_1$ prevents any issues with the mathematical singularities at the origin.  The number of oscillations of the wave-field near $r=0$ increases with the order $\nu$ of the mode.  Even for modes with real fractional orders that result for anisotropic fibers, there is a comparatively larger confinement  of the higher-order $m$ modes (Fig.~\ref{fg:schematic}). The solutions for some higher order modes in anisotropic fibers for fractional orders and imaginary orders are shown in Fig.~\ref{fg:schematic}, where the solutions for modes with integral order in a positive index waveguide are also shown for comparison.

The origin of this spreading out of the fields of the modes with large $m$ well into the interior of the waveguide lies in the dispersion of the wave in a hyperbolic medium with cylindrical symmetry.  Consider at very small length scales, when a wave propagating in the $(r,\phi)$ plane  would essential look like a plane wave, so that the local wave-vector can be considered to be along the ray. If $(k_r,k_\phi)$ be locally the wave-vector, the phase shift for an infinitesimal displacement would be $\exp[i (k_r \delta r + k_\phi r\delta\phi)]$. Demanding a single valued function constrains that $k_\phi = m /r$, where $m$ is an integer. Under these circumstances (in analogy to a wave in an anisotropic medium in flat rectangular space), locally we have 
\begin{equation}
\label{eq:dispersion}
\frac{k_r^2}{\varepsilon_\phi} + \frac{m^2}{r^2\varepsilon_r} = \frac{\omega^2}{c^2}.
\end{equation}
In a normal positive medium with $\varepsilon_r > 0$ and $\varepsilon_\phi >0$, as the wave moves towards the origin, $r\rightarrow 0$ and $k_\phi \rightarrow \infty $, and $k_r$ becomes imaginary. Thus, the wave decays exponentially towards the origin for the large $m$ whispering gallery modes and avoids the region near the origin. On the contrary, in an hyperbolic medium with $\varepsilon_r \cdot \varepsilon_\phi <0$, $k_r$ remains real and large for $k_\phi \rightarrow \infty$ (large $m$), and the wave can effectively propagate in the region near the origin with the only difference that the fields oscillate more and more rapidly along $r$ closer to the origin. This is due to the fact that propagation of  higher-order modes that can be supported in media with hyperbolic dispersion as seen earlier in the case of the hyperlens~\cite{hyperlens_OE}. In that work the higher-order modes carrying spatially sub-wavelength (rapidly varying) image features could be propagated across the radial direction and imaged out into far-field radiating modes~\cite{hyper_lens}.   For  example,  in Eq. (\ref{eqn_TM_dispersion}), if $\varepsilon_r<0$ and $\varepsilon_z>0$,  then $k_r$ increases without bound along with $\beta$. Thus, very large oscillatory field variations in the radial direction can be supported close the centre of the fibre  and can have large modal volumes.  

\begin{figure}
\begin{center}
\includegraphics[width=70mm]{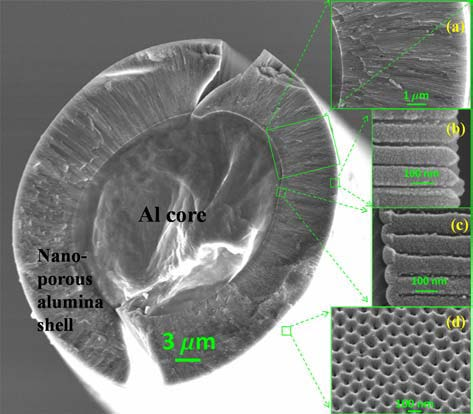}
\end{center}
\caption{ \label{fg:NAA_mutube} Scanning electron micrographs of the nanoporous alumina microtube. Note the presence of the radially oriented non-branching nanopores. The nanoporous outer surface and the impermeable barrier oxide layer at the inner tubular surface are shown in the insets. The brittle alumina microtube cracks when cleaved for SEM imaging.}
\end{figure}

Whereas the anisotropy described above may be realized in microwave waveguides using thin-wire metamaterials~\cite{hcirfiber,unpub}, the task of practically fabricating a metamaterial fiber for optical/NIR frequencies with nanoscale features is daunting.  However, it was determined that a wire mesh metamaterial in the cylindrical geometry could be obtained by anodization  electrodeposition methods.  Starting with a high purity (99.999\%)  aluminum  wire that was cleaned and electropolished,  a two-step anodization~\cite{Masuda} was carried out at 40V in 0.3 M oxalic acid  to obtain a nanoporous alumina layer of several micrometers to several tens of micrometers thickness.  The nanoporous alumina consists of an inner impermeable alumina layer near the center from which nanopores radially emanate and terminate on a nanoporous surface formed by the nanopores as shown in Fig.~\ref{fg:NAA_mutube}.  The nanopores may now filled with a plasmonic metal like silver, gold or nickel by electrodeposition techniques~\cite{wiremtm}.  This results in an extremely large anisotropy in the waveguide. The aluminum wire at the center may be retained or etched away using CuCl$_2$. If retained, it effectively forms a PEC at the inner surface ($R_1$), while a simple thick coating of aluminum on the outer nanoporous surface will also provide an effective PEC there ($R_2$). This yields the exact physical manifestation of the coaxial metamaterial fiber discussed above.  The important structural parameters of the nanoporous alumina fiber are the thickness of the porous alumina layer, the inner core diameter and interpore separation and pore sizes. These parameters are flexibly controlled by choice of the thickness of the initial aluminum wire, acid environment and  anodization voltage and the times for electropolishing,  anodization and etching.  Microtubes have been fabricated with lengths of up to 5 cm with alumina thicknesses ranging from 2 $\mu$m to 80 $\mu$m, and inner and outer diameters ranging from 3 $\mu$m to 40 $\mu$m and 100 $\mu$m, respectively.

It is well known that a metamaterial of unidirectionally oriented wires in rectangular coordinates constitutes an anisotropic hyperbolic dispersive medium~\cite{sridhar_APL,wiremtm,homogenization}.  Using a mapping technique based on transformation optics~\cite{Ward,Pendry,PendrySAR}, we show that an effective anisotropic dielectric tensor  can be  obtained for the medium with radially oriented wires / pores in the curvilinear geometry as well. The cylindrical shell can be transformed into a rectangular slab by the transformation
\begin{equation}
\tilde{x}=\ln r, \quad \tilde{y}=\phi, \quad \tilde{z}=z,
\label{seq:coortrans}
\end{equation}
and in the new coordinate frame, the material parameters and fields can be written as
\begin{equation}
 \tilde{\varepsilon_j}=\varepsilon_j \frac{S_1 S_2 S_3}{{S_j}^2}, \quad \tilde{\mu_j}=\mu_j \frac{S_1 S_2 S_3}{{S_j}^2}, \quad  \tilde{E_j}=S_j E_j, \quad \tilde{H_j}=S_j H_j,
 \label{seq:epmugen}
\end{equation}
\begin{equation}
 S_j^2=\left( \partial r \over \partial \tilde{q_j} \right)^2+\left(r \partial \phi \over \partial \tilde{q_j} \right)^2+\left( \partial z \over \partial \tilde{q_j} \right)^2,
 \label{seq:s}
\end{equation}
where  tilde ($~\tilde{}~$)  represents the new transformed frame. Now using Eqs. (\ref{seq:coortrans})--(\ref{seq:s}) the material parameters in the new frame for the embedded material ($i$) and alumina host material ($h$) within their respective regions are
\begin{equation}
 \tilde{\varepsilon}_{\tilde{x},s}=\varepsilon_{r,s}, \tilde{\varepsilon}_{\tilde{y},s}=\varepsilon_{\phi,s},  \tilde{\varepsilon}_{\tilde{z},s}=e^{2 \tilde{x}}\varepsilon_{z,s}; \quad \tilde{\mu}_{\tilde{x},s}=\mu_{r,s},  \tilde{\mu}_{\tilde{y},s}=\mu_{\phi,s},  \tilde{\mu}_{\tilde{z},s}=e^{2 \tilde{x}}\mu_{z,s},
 \label{seq:epi}
\end{equation}
where $s=i, h$.
\begin{figure}
\begin{center}
\includegraphics[width=100mm]{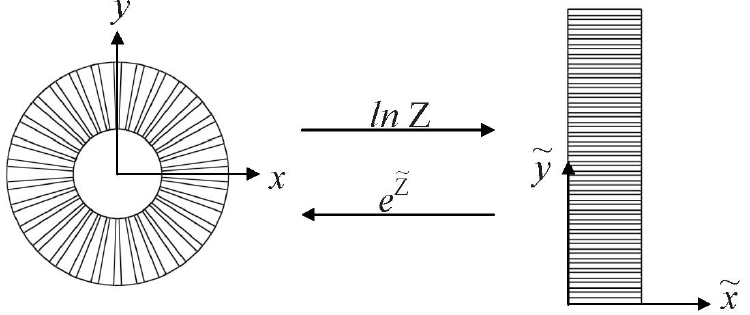}
\end{center}
\caption{ \label{fg:geometry_mapping} A cylindrical shell is mapped to a flat slab. The nanopores along the radial direction in shell mapped along the $\tilde{x}$-direction in the new frame.}
\end{figure}

In this manner, we transform the cylindrical shell with radially emanating  pores/nanorods into a slab with columns along the $\tilde{x}$ direction (see Fig.~\ref{fg:geometry_mapping}). The only difference from a medium of unidirectional nano-column is that both the background material of the slab and the columns are both anisotropic and inhomogenous (along $\tilde{z}$).
Such a structure can be easily homogenized by noting that the principal axis of anisotropy coincides with the column axes and we have only diagonal anisotropy.

Now we will proceed to homogenise the structure in the new geometry, and then  transform the obtained effective-medium parameters back to the cylindrical geometry. This will be valid at all points, except at the origin in the cylindrical geometry where there is a singularity, which is excluded in the homogenization. The effective permittivity along the $\tilde{x}$-direction with the help of Eq. (\ref{seq:epi}) can be given by the volume average~\cite{Chakrabarti,homogenization}
\begin{equation}
  \varepsilon_{\tilde{x}}^{eff}=f \varepsilon_{r,i}+(1-f)\varepsilon_{r,h},                               
  \label{seq:epxteff2}
\end{equation}
where $f$ is the fill fraction of the inclusion. In the transverse direction, the effective permittivity can be approximately obtained by the Maxwell-Garnett homogenization procedure. We would need the polarizabilities of anisotropic cylinders embedded in an anisotropic medium.  The polarisability of an
 anisotropic prolate spheroid in a isotropic medium is well known~\cite{Bohren} and the homogenization of anisotropic ellipsoids imbedded in a anisotropic medium has been also been obtained~\cite{Jones}. In our case, as the inclusion is cylindrical with very large aspect ratio, therefore, it can be approximated as a prolate spheroid with extreme eccentricity (like a long needle).  The homogenized effective-medium parameters are obtained as
\begin{equation}
  \varepsilon_{\tilde{y}}^{eff} = \frac{(1+f) \varepsilon_{\phi,i} \varepsilon_{\phi,h} + (1-f) \varepsilon_{\phi,h}^2}{(1-f) \varepsilon_{\phi,i} + (1+f) \varepsilon_{\phi,h}}     ,                       
  \label{seq:epyteff2}
\end{equation}
and 
\begin{equation}
  \varepsilon_{\tilde{z}}^{eff} = e^{2\tilde{x}}\frac{(1+f) \varepsilon_{z,i} \varepsilon_{z,h} + (1-f) \varepsilon_{z,h}^2}{(1-f) \varepsilon_{z,i} + (1+f) \varepsilon_{z,h}}.                            
  \label{seq:epzteff2}
\end{equation}
The additional exponential factor in the $\varepsilon_{\tilde{z}}^{eff}$ results in a biaxial and inhomogeneous medium in the Cartesian coordinate frame of rectangular slab. Thus, it is different from a flat nanoporous  anodic alumina template. 

Now we will map back these parameters into the original cylindrical geometry and obtain 
\begin{equation}
  \varepsilon_{r}^{eff}=f \varepsilon_{r,i}+(1-f)\varepsilon_{r,h},                              
  \label{seq:epxeff}
\end{equation}
\begin{equation}
  \varepsilon_{\phi}^{eff} = \frac{(1+f) \varepsilon_{\phi,i} \varepsilon_{\phi,h} + (1-f) \varepsilon_{\phi,h}^2}{(1-f) \varepsilon_{\phi,i} + (1+f) \varepsilon_{\phi,h}},                            
  \label{seq:epyeff}
\end{equation}
and
\begin{equation}
  \varepsilon_{z}^{eff} = \frac{(1+f) \varepsilon_{z,i} \varepsilon_{z,h} + (1-f) \varepsilon_{z,h}^2}{(1-f) \varepsilon_{z,i} + (1+f) \varepsilon_{z,h}}.                            
  \label{seq:epzeff}
\end{equation}
Note that the exponential factor of $e^{2\tilde{x}}$ cancels  in the  inverse mapping process. Since in the cylindrical geometry the individual materials are isotropic, the effective permittivity components $\varepsilon_{\phi}^{eff}$ and $\varepsilon_{z}^{eff}$ are equal in this approximation. Incidentally, we obtain an identical result by just applying this process directly in the curvilinear geometry. This is the due to the large aspect ratio of the inclusions  and the electrically small diameter of the nanopores aligned along the $r$-direction. We also note  that cylindrically symmetric metamaterials have been also been homogenised using such geometric transformations~\cite{Nicolet,Zollabook}, including  for the design of anisotropic invisibility cloaks~\cite{cyl_homogenized_cloak}.

\begin{figure}
\begin{center}
\includegraphics[width=110mm]{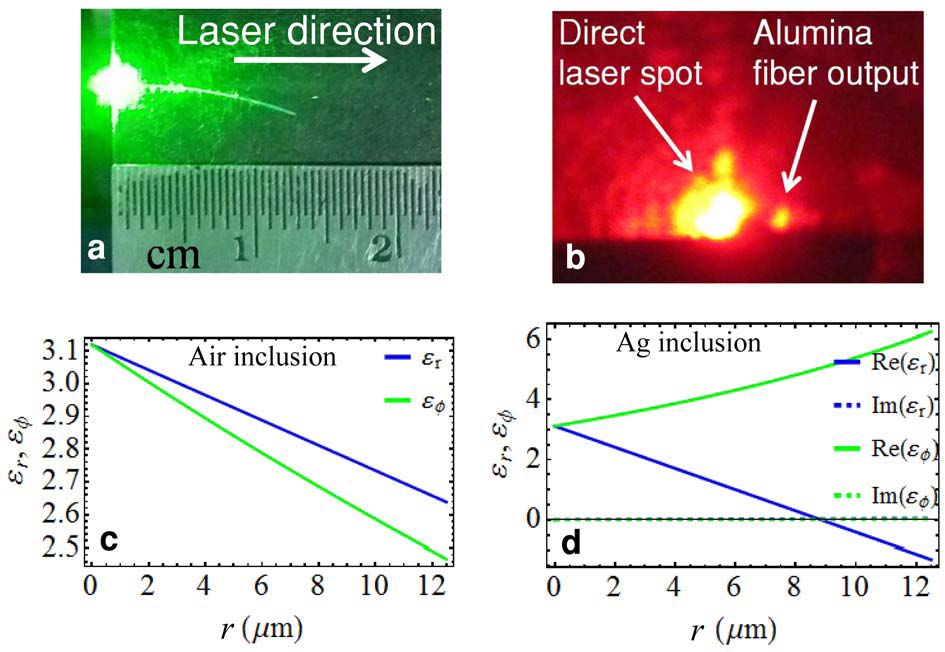}
\end{center}
\caption{ \label{fg:fiber_lightguide} Top: Picture of light ($\lambda=532$ nm) propagating across a bent nanoporous alumina fiber with an aluminum core (Aluminium core diameter- 10 $\mu$m, nanoporous alumina shell diameter- 80 $\mu$m, length- 1.3 cm, nanopore diameter is 30 nm and nanopore periodicity is 100 nm at outer surface). The output from the fiber at $\lambda=633$ nm is shown on the right. Plot of the variation of effective dielectric permittivity components in the Maxwell-Garnet approximation with the radial distance in a nanoporous alumina microtube for air inclusion (Bottom-left) and when the nanopores are filled with silver for nanopore radius $q=25$ nm at outer surface and $f=0.23$(bottom-right).}
\end{figure}

In this picture, only the areal fill fraction ($f$) of the pore/nanorod determines the effective medium properties and the anisotropy results in $\varepsilon_r \ne \varepsilon_\phi$ and $\varepsilon_z  \simeq \varepsilon_\phi$. From the SEM images of several samples, we find that the nanopore cross-section approximately reduces linearly with the radial distance in alumina microtubes  anodized at constant voltage, and the areal fill fraction changes with the radial distance as  $f = (2\pi q^2 r)/(\sqrt{3}d^2 R))$,
where $2q$ is the pore diameter and $d$ is the pore separation at some given radius $R$. Typically, $d=100$ nm at the surface of the aluminium wire where anodization begins when an anodization voltage of 40 V is used at 0$^{\circ}$ C in oxalic acid. Similarly, the pore diameter $2q$ is 30 nm under these circumstances. The pore diameter can be flexibly increased to any any value up to about 90 nm by pore widening in a solvent such as phosphoric acid.  Fig.~\ref{fg:fiber_lightguide} shows the variation in the anisotropic material parameters with radial distance obtained for air nano pores in nanoporous alumina and silver nanorods in nanoporous alumina where a Drude dielectric permittivity for silver was used as $\varepsilon_{Ag} = \varepsilon_\infty  - \omega_p^2/[\omega(\omega+i\gamma)]$ with $\varepsilon_\infty =5.7$, $\omega_p=9.2$eV and $\gamma=0.021$eV~\cite{shalaevbook}. For host alumina the data have been taken from~\cite{handbook}. The imaginary part of the effective medium parameters for the silver filled nanoporous structure is also small due to the small fill fractions of silver.  Due to the inhomogeneous radial variation of the pore diameters, these fibers are actually spatially inhomogenous, which will affect the nature of the modes. Typically, nanowires made of plasmonic materials that have $\mathrm{Re}(\varepsilon) <0$ results in an extreme effective anisotropy with $\varepsilon_r <0$ and $\varepsilon_\phi >0$. While anisotropy itself makes it possible to have modes with imaginary orders, the indefinite permittivity with $\varepsilon_r < \varepsilon_\phi$ (as $\varepsilon_r < 0$ and $\varepsilon_\phi > 0$) that results when plasmonic nanorods fill the nanopores makes the conditions for modes with imaginary orders immediately accessible.  

The guidance of light through a bent section of a nanoporous anisotropic fiber and a aluminium core is shown in Fig.~\ref{fg:fiber_lightguide}. The scattering from inhomogeneities and other defects is strong, but the evidence for light confinement and guidance is clear. Due to the large levels of scattering, the modes are all coupled, and we have not been able to image the mode structure of the propagation modes. The propagation loss is also very large due to the scattering.  Although the nanoporous alumina material obtained here might not possess optimal optical properties, improved processing will result in lower scattering losses. The nanostructure may be converted into silicate glasses through chemical processes or sol-gel methods similar to those in~\cite{Masuda}. The nanostructured nature of these fibers always contributes some scattering loss; nevertheless, it is clear that that low-loss anisotropic optical fibers can be obtained. 
It is also important to realize that the generic nature of these modal solutions are not limited by the application of PEC boundaries. The only restriction would be for the nanoporous alumina regions to have a larger dielectric permittivity than the surrounding medium. Having dielectric boundaries, for example by having a hollow core and no aluminum coating on the outer surface, causes the waveguide to be inhomogeneous and support hybrid (the so-called HE and EH) modes.  Fibers with hollow cores are important for various issues of confinement of light as in hollow core photonic crystal fibers. The special case in which the core is hollow but the PEC boundary condition at $r=R_2$ is maintained is, in fact, an anisotropic metamaterial-lined circular waveguide. When the metamaterial liner exhibits certain properties, e.g. a permittivity tensor containing negative and near-zero elements, these waveguides permit propagation well below their natural cutoff frequencies~\cite{belowcutoff}. This has important applications in the miniaturization of microwave waveguide systems requiring access to the interior volume such as in fluid heating, electron-beam propagation, and  traveling-wave magnetic-resonance imaging. As open-ended waveguide probe antennas, these structures may enable subwavelength spatial-resolution measurements with high efficiency in near-field antenna characterization~\cite{unpub}. We also note the case of TM modes with imaginary orders that have no cutoff frequency when $\varepsilon_z <0$ and $\varepsilon_r >0$,  as then $\beta^2 = \varepsilon_r \omega^2/c^2 + \varepsilon_r/ \vert \varepsilon_z\vert k_r^2$ that are distinct from the quasi-static TEM mode.

We now discuss a sample of the astonishing range of possibilities that arise with anisotropic metamaterial optical fibers.  The Bessel modes with imaginary orders have enormous implications for coupling to the near-field modes of small sources as they have fast varying fields along both the radial and azimuthal directions localized near the center of the fiber. The subwavelength image resolution possible with near-field imaging by a hyperlens is also due to this effect~\cite{hyperlens,hyperlens_OE}.  Usually, one is able to couple light to the whispering gallery modes only though the cladding due to the localization of these modes at the edges. The Bessel modes with imaginary orders can enable a butt-coupling to the high order $m$ modes to couple to the near-field modes of a radiating source.  Further, these Bessel modes with imaginary orders occupy the entire cross section of the fibre and have implications for the amplification of these modes in the presence of gain media without spatial hole burning effects. The presence of plasmonic nanorods inside the medium results in large local electromagnetic fields that are crucial for nonlinear interactions. The guidance of fields within the fiber with large nonlinear interactions has immense implications for effects like super-continuum generation due to dispersion of ultra-short pulses.  The enormous surface areas available within the nanopores for adsorbing molecules coupled with the fiber geometry and resonant modes makes this system highly suitable for sensor applications. In case of hollow core fibers, liquids can be flown through the central microtube and the enhanced fields of the higher order $m$ modes localized near the center could be a very useful for detecting molecules in the liquids.

A discussion of dissipation and its effects on the modes with imaginary orders is also imperative. Alumina is an excellent optical material for visible and infrared frequencies with low absorption. However, anodized alumina can have larger absorption in general~\cite{absor}. The incorporation of silver will also cause more dissipation, resulting in a complex propagation constant $\beta$. Using the solutions for modes for homogenised fibers corresponding to silver-filled nanoporous alumina micro tubes, the ratio of the imaginary and real parts of $\beta$ was on the order of $10^{-3}$ for TM$_{1,1}$ modes ($f=0.08$) with complex order.  Thus, while the fibers discussed here are clearly not meant for traditional transmission applications, they have immense potential for several other purposes such as coupling the near-field into a waveguide. 

We have demonstrated here a  new paradigm of metamaterial fibers and circular waveguides.  These anisotropic metamaterial fibers support exotic modes described by Bessel and Neumann functions of fractional and imaginary orders,  a rare physical application of these mathematical objects.  Many of the properties presented here are generic to circular waveguides with anisotropic metamaterial fillings and have wide ranging applications across the electromagnetic spectrum from the radio frequencies to optical frequencies. Nanoporous alumina microtubes  with or without a central hollow core are presented as a physical manifestation of such an anisotropic metamaterial optical fiber.

\section*{Acknowledgements}
DP thanks the CSIR India for a fellowship, SAR acknowledges funding from the DST,India under the project no. DST/SJF/PSA-01/2011-2012. JP and AKI acknowledge funding from the Natural Sciences and Engineering Research Council of Canada.



\begin{thebibliography}{99}
\bibitem{Russell}P. Russell, ``Photonic crystal fibers,'' Science {\bf 299}, 358-362 (2003).
\bibitem{PCFcoil}C. F. Fan, C. L. Chiang, and C. P. Yu, ``Birefringent photonic crystal fiber coils and their application to transverse displacement sensing,'' Opt. Express {\bf 19}, 19948-19954 (2011).
\bibitem{sarbook}S. A. Ramakrishna and T.M. Grzegorczyk, {\it Physics and Applications of Negative Refractive Index Materials}, (CRC Press, 2009).
\bibitem{sridhar_APL} B. D. F. Casse, W. T. Lu, Y. J. Huang, E. Gultepe, L. Menon and S. Sridhar, ``Super-resolution imaging using a three-dimensional metamaterials nanolens'', Appl. Phys. Lett. {\bf 96}, 023114 (2010)
\bibitem{unmu}S. Hrabar, J. Bartolic and Z. Sipus, ``Waveguide miniaturization using uniaxial negative permeability metamaterial,'' IEEE Trans. Antennas Propag. {\bf 53}, 110-119 (2005).
\bibitem{belowcutoff}J.G. Pollock and A.K. Iyer, ``Below-cutoff propagation in metamaterial-lined circular waveguides,'' IEEE Trans. Microw. Theory Techn. {\bf 61}, 3169-3178 (2013).
\bibitem{novitsky_barkovsky}A. V. Novitsky, and L. M. Barkovsky, ``Guided modes in negative-refractive-index fibres,'' J. Opt. A: Pure Appl. Opt. {\bf 7}, S51–S56 (2005).
\bibitem{novitsky}A. V. Novitsky, ``Negative-refractive-index fibres:TEM modes,'' J. Opt. A: Pure Appl. Opt. {\bf 8}, 864–866 (2006).
\bibitem{cory_blum}H. Cory, and T. Blum, ``Surface-wave propagation along a metamaterial cylindrical guide,'' Microwave Opt. Tech. Lett. {\bf 44}, 31–35 (2005).
\bibitem{IIT_kgp}B. Ghosh, and A. B. Kakade, ``Guided modes in a metamaterial-filled circular waveguide,'' Electromagnetics A {\bf 32}, 465–480 (2012).
\bibitem{Smith}E. J. Smith, Z. Liu, Y.Mei, and O.G. Schmidt, ``Combined surface plasmon and classical waveguiding through metamaterial fiber design,'' Nano Lett. {\bf 10}, 1-5 (2010).
\bibitem{emode}M. Yan, N. A.  Mortensen and M. Qui, ``Engineering modes in optical fibers with metamaterial,'' Front. Optoelectron. China {\bf 2}, 153-158 (2009).
\bibitem{atakaramians_JOSAB_2012}S. Atakaramians, A. Argyros, S. C. Fleming, and B. T. Kuhlmey, ``Hollow-core waveguides with uniaxial metamaterial cladding: modal equations and guidance conditions,'' J. Opt. Soc. Am. B {\bf 29}, 2462-2477 (2012).
\bibitem{atakaramians_JOSAB_2013}S. Atakaramians, A. Argyros, S. C. Fleming, and B. T. Kuhlmey, ``Hollow-core uniaxial metamaterial clad fibers with dispersive metamaterials,'' J. Opt. Soc. Am. B {\bf 30}, 851-867 (2013).
\bibitem{nsingh_OE2012}N. Singh, A. Tuniz, R. Lwin, S. Atakaramians, A. Argyros, S. C. Fleming, and B. T. Kuhlmey, ``Fiber-drawn double split ring resonators in the terahertz range,'' Opt. Mat. Express {\bf 2}, 1254-1259 (2012).
\bibitem{Masuda}H. Masuda and K. Fukuda, ``Ordered metal nanohole arrays made by a two-step replication of honeycomb structure of anodic alumina,'' Science {\bf 268}, 1466-1468 (1995).
\bibitem{Dunster}T.M. Dunster, ``Bessel functions of purely imaginary order, with an application to second-order linear differential equations having a large parameter,'' SIAM J. Math. Anal. {\bf 21}, 995-1018 (1990).
\bibitem{Chapman}C. J. Chapman, ``The asymptotic theory of dispersion relations containing Bessel functions of imaginary order,'' Proc. R. Soc. A {\bf 468}, 4008-4023 (2012).
\bibitem{Grimshaw}R. H. J. Grimshaw, K. R. Khusnutdinova, L. A. Ostrovsky, ``The effect of a depth-dependent bubble distribution on the normal modes of internal waves: quasistatic approximation,'' Eur. J. Mech. B/Fluids {\bf 27}, 24-41 (2008).
\bibitem{hyperlens_OE}Z. Jacob, L. V. Alekseyev, and E. Narimanov, ``Optical hyperlens: far-field imaging beyond the diffraction limit,'' Opt. Express {\bf 14}, 8247-8256 (2006).
\bibitem{hyper_lens}Z. Liu, H. Lee, Y. Xiong, C. Sun, X. Zhang, ``Far-field optical hyperlens magnifying sub-diffraction-Limited objects,'' Science {\bf 315}, 1686-1686 (2007).
\bibitem{hcirfiber}M. Yan, N. A.  Mortensen, ``Hollow-core infrared fiber incorporating metal-wire metamaterial,'' Opt. Express {\bf 17}, 14851-14864 (2009).
\bibitem{unpub} J. G. Pollock and A. K. Iyer, "Miniaturized circular-waveguide probe antennas using metamaterial liners," IEEE Trans. Antennas Propagat., {\bf 63} , 428-433 (2015).
\bibitem{wiremtm}C.R. Simovsky, P.A. Belov, A.V. Atrashenko, and Yu. S. Kivshar, ``Wire metamaterials: physics and applications,'' Adv. Mat. {\bf 24}, 4229-4248 (2012).
\bibitem{Ward}A. J. Ward, and J. B. Pendry, ``Refraction and geometry in Maxwell's equations,'' J. Mod. Opt. {\bf 43}, 773-793 (1996).
\bibitem{Pendry}J. B. Pendry, ``Perfect cylindrical lenses,'' Opt. Express {\bf 11}, 755-760 (2003).
\bibitem{PendrySAR}J. B. Pendry, and S. A. Ramakrishna ``Focusing light using negative refraction,'' J. Phys.: Condens. Matter {\bf 15}, 6345–6364 (2003).
\bibitem{Nicolet}A. Nicolet, F. Zolla, and Y. Ould Agha, ``Geometrical transformations and equivalent materials in computational electromagnetism,'' COMPEL {\bf 27}, 806-819 (2008).
\bibitem{Zollabook}F. Zolla, G. Renversez, A. Nicolet , B. Kuhlmey , S. Guenneau , D. Felbacq , and A. Argyros, S. Leon-Saval, {\it Foundations of Photonic Crystal Fibres}, (Imperial College Press, 2008).
\bibitem{Chakrabarti}S. Chakrabarti, S. A. Ramakrishna, and H. Wanare, ``Switching a plasmalike metamaterial via embedded resonant atoms exhibiting electromagnetically induced transparency,'' Opt. Letters {\bf 34}, 3728-3730 (2009).
\bibitem{homogenization}J. Elser, R. Wangberg, V. A. Podolskiy, and E. E. Narimanov, ``Nanowire metamaterials with extreme optical anisotropy,'' Appl. Phys. Lett. {\bf 89}, 261101-261103 (2006).
\bibitem{Bohren}C. F. Bohren, and D. R. Huffman, {\it Absorption and scattering of light by small particles}, (John Wiley, New York, USA, 1983).
\bibitem{Jones}R. C. Jones, ``A generalization of the dielectric ellipsoid problem,'' Phys. Rev. {\bf 68}, 93-96 (1945).
\bibitem{cyl_homogenized_cloak}M. Farhat, S. Guenneau, A.B. Movchan, and S. Enoch, ``Achieving invisibility over a finite range of frequencies,'' Opt. Express {\bf 14}, 5658-5661 (2008).
\bibitem{shalaevbook}W. Cai, and V. Shalaev, {\it Optical Metamaterials: Fundamentals and Applications}, (Springer, 2010).
\bibitem{handbook}M. J. Weber, {\it Handbook of Optical Materials}, (CRC Press, 2003).
\bibitem{hyperlens}Z. Liu, H. Lee, Y. Xiong, C. Sun, and X. Zhang, ``Far-field optical hyperlens magnifying sub-diffraction-limited objects,'' Science {\bf 315}, 1686 (2007).
\bibitem{absor}W. J. Zheng, G. T. Fei, B. Wang, and L. D. Zhang, ``Modulation of transmission spectra of anodized alumina membrane distributed Bragg reflector by controlling anodization temperature,'' Nanoscale Res Lett {\bf 4}, 665–667 (2009).
\end{thebibliography}
\end{document}